\documentclass{svproc}

\usepackage[greek, english]{babel}
\usepackage{bbm} 
\usepackage{amsmath, amsfonts, amssymb, mathtools}    
\usepackage{color}      
\usepackage{subfigure}  
\usepackage{epstopdf}
\usepackage{hyperref}   
\usepackage{tabularx}
\usepackage{dcolumn}

\usepackage{physics}
\usepackage{slashed} 
\usepackage{tensor}

\newcommand{\moye}[1]{\langle #1 \rangle}
\def\man{\mathcal{M}}

\def\O{\mathcal{O}}

\def\Psib{\bar{\Psi}}
\def\dag{\dagger}
\def\pa{\partial}

\def\sl{\slashed}

\DeclareMathOperator\Det{Det}

\def\R{\ensuremath{\mathbb{R}}}

\usepackage{cancel}

\def\lb{\left(}
\def\rb{\right)}

\def\l.{\left.}
\def\r.{\right.}

\def\beq{\begin{equation}}
\def\eeq{\end{equation}}
\def\bsp{\begin{split}}
\def\esp{\end{split}}
\def\bea{\begin{eqnarray}}
\def\eea{\end{eqnarray}}
\def\beano{\begin{eqnarray*}}
\def\eeano{\end{eqnarray*}}

\newcommand{\eqn}[1]{\begin{align}#1\end{align}}

\newcommand{\pmatr}[1]{\begin{pmatrix}#1\end{pmatrix}}

\definecolor{gris}{rgb}{0.4,0.4,0.4}

\usepackage{bm}

\DeclareMathOperator\mQED{QED}
\DeclareMathOperator\QEDt{QED_3}
\DeclareMathOperator\QEDtGN{QED_3-GN}
\DeclareMathOperator\QEDtcHGN{QED_3-cHGN}

\DeclareMathOperator\cHGN{cHGN}

\DeclareMathOperator\U{U}
\DeclareMathOperator\SU{SU}

\def\nf{N}
\def\d{\downarrow}
\def\u{\uparrow}
\def\l{\leftarrow}
\def\r{\rightarrow}

\def\lag{\mathcal{L}}

\usepackage{amsmath, amsfonts,amssymb,mathtools}
\usepackage{bbm}
\usepackage{graphicx}
\usepackage{epstopdf}

\usepackage{array}

\usepackage{ytableau}
\usepackage{youngtab}

\newcommand{\svdots}{%
  \vbox{
     \baselineskip 2pt \lineskiplimit 0pt
    \hbox {.}\hbox {.}\hbox {.}\kern-0.75pt
  }%
}
\newcommand{\sdots}{%
  \vbox{
     \baselineskip 2pt \lineskiplimit 0pt
     \dots \kern-0.75pt
  }%
}
\ytableausetup{smalltableaux}
\Yboxdim{8pt}

\usepackage{lipsum}

\def\veps{\varepsilon}

\usepackage{url}

\usepackage{cite}

\begin{document}
\mainmatter              
\title{Monopole operators and their symmetries in QED$_3$-Gross-Neveu models}
\titlerunning{Monopole operators and their symmetries in QED3-Gross-Neveu models}  
%
\author{\'Eric Dupuis\inst{1}, M. B. Paranjape\inst{1,2}, William Witczak-Krempa\inst{1,2}}
\authorrunning{\'Eric Dupuis et al.} 
%
\tocauthor{\'Eric Dupuis, M.B. Paranjape, William Witczak-Krempa}
\institute{D\'epartement de physique, Universit\'e de Montr\'eal, Montr\'eal (Qu\'ebec), H3C 3J7, Canada\\
\email{eric.dupuis.1@umontreal.ca} 
\and
Centre de Recherches Math\'ematiques, Universit\'e de Montr\'eal; P.O. Box 6128, Centre-ville Station; Montr\'eal (Qu\'ebec), H3C 3J7, Canada}

\maketitle              

\begin{abstract}
Monopole operators are topological disorder operators in $2+1$ dimensional compact gauge field theories appearing notably in quantum magnets with fractionalized excitations. For example, their proliferation in a spin-1/2 kagome Heisenberg antiferromagnet triggers a quantum phase transition from a Dirac spin liquid phase to an antiferromagnet. The quantum critical point (QCP) for this transition is described by a conformal field theory :  Compact  quantum electrodynamics (QED$_3$) with a fermionic self-interaction, a type of QED$_3$-Gross-Neveu model. We obtain the scaling dimensions of monopole operators at the QCP using a state-operator correspondence and a large-$N_f$ expansion, where $2 N_f$ is the number of fermion flavors.   We characterize the hierarchy of monopole operators at this $\SU(2) \times \SU(N_f)$ symmetric  QCP. 

\end{abstract}
\section{Confinement of a Dirac spin liquid}
An abelian gauge theory consists of the Maxwell term for the gauge field $a_\mu$ and potentially some  matter coupled  through a $\U(1)$ charge
\eqn{
\lag  = \dfrac{1}{2 e^2} \lb \epsilon_{\mu \nu \rho} \pa_\nu a_\rho \rb^2 + \text{Matter charged under $\U(1)$}\,.
}
In $2+1$ dimensions, the magnetic current  is constructed by contracting the {rank-$3$}  antisymmetric tensor with the field strength $j^\mu_{\rm top} = \frac{1}{2\pi} \epsilon^{\mu \nu \rho}  \pa_\nu a_\rho$. The current conservation then expected  is  violated when regularizing the theory on the lattice. Indeed, the gauge field  $a_{\mu}$ then becomes periodic, taking values in the compact $\U(1)$ gauge group.  This implies the existence  of monopole operators $\man_q^\dag$ which create gauge field configurations $A^q$   that may be written as\footnote{In vector notation using spherical coordinates  on Euclidean spacetime $\R^3$, it would be written as $A^q_\mu = q(1-\cos \theta)/(r \sin \theta) \delta_\mu^{\phi}$.}
\eqn{
A^{q} = q (1 - \cos \theta ) \dd \phi \,, \label{eq:Aq}
}
where the magnetic charge $q$ is half-quantized to respect the Dirac condition. In turn, this implies $2\pi$ quantization of the magnetic flux  $\Phi = \int_{S^2} \dd A^q =  4 \pi q$. 

Monopole operators may render a gauge theory unstable. In the compact pure $\U(1)$ gauge theory in $2+1$ dimensions, monopole operators are relevant and condense. This leads to confinement and to the emergence of a mass gap \cite{polyakov_compact_1975, polyakov_quark_1977}. Adding massless matter may, however, stabilize the gauge theory. For a large number $N$ of massless matter flavors, the monopole two-point function is 
\eqn{
\moye{\man_{q}(x)\man_{q}^{\dagger}(0)} \sim x^{-2\Delta_{\man_q}} \approx x^{-2  N (\dots)}\,, \quad N \gg 1\,,
}
where $\Delta_{\man_q}$ is the scaling dimension of the monopole operator, and the ellipses~$(\dots)$ denote  a number of order $1$ \cite{borokhov_topological_2003, metlitski_monopoles_2008}. Monopoles are thus suppressed for a sufficient number of flavors $N$.  Interestingly,  a confinement-deconfinement transition  can therefore be achieved by tuning an interaction which gaps the massless matter and removes its stabilizing screening effect.

The stability of compact  gauge theories plays a key role in  strongly correlated systems where fractionalized quasiparticles  and gauge excitations emerge.  In particular, certain frustrated 2D quantum magnets may be described at low energy by a Dirac spin liquid (DSL). This  is a version of quantum electrodynamics in three dimensions ($\mQED_3$),
\eqn{
\lag_{\mQED_3} =   - \Psib \sl{D}_a \Psi +  \frac{1}{2e^2} \lb \epsilon_{\mu \nu \rho} \pa_\nu a_{\rho} \rb^2 \,,
\label{eq:QED3}
}
with  $2\nf$ flavors of two-component gapless Dirac fermions    $\Psi = \pmatr{\psi_1\,,&\dots&, \psi_{2\nf}}^\intercal$. The gauge covariant derivative is defined as $\sl{D}_a  = \gamma_\mu \lb \pa_\mu - i  a_\mu \rb$  where $\gamma_\mu$ are the Pauli matrices.
The  $2\nf$ fermion flavors can represent the two magnetic spins and $\nf$ Dirac nodes in momentum space, typically two as well. In particular, many numerical studies suggest that a DSL with  $\nf=2$ Dirac cones may describe the ground state of the spin-1/2 kagome Heisenberg antiferromagnet \cite{ran_projected-wave-function_2007, sindzingre_low-energy_2009,  iqbal_gapless_2013, iqbal_spin_2015,  he_signatures_2017, zhu_entanglement_2018}. 

The stability of the DSL then hinges on the irrelevance of monopole excitations allowed by the lattice. Whether it is stable or not at $\nf=2$ is  still an ongoing question \cite{pufu_anomalous_2014, meng_monte_2019, song_unifying_2019}. Here, we suppose a stable DSL and we focus on the possible confinement-deconfinement transition. In this context, the transition may be driven by a Gross-Neveu type interaction $\delta \lag = -\frac{h^2}{2}  (\Psib T^a \Psi)^2$, where $T^a$ is a generator of the flavor symmetry group $\SU(2 \nf)$. For a sufficiently strong coupling, the corresponding fermion bilinear  condenses $\moye{ \Psib T^a \Psi } \neq 0$, allowing  monopoles to proliferate. Importantly, different types of monopole operators exist in $\mQED_3$. This is because  half of the $4 q \nf$ fermion zero modes on a monopole must be filled to minimize its scaling dimension and yield a vanishing fermion number \cite{atiyah_index_1963, borokhov_topological_2003}.   Therefore, the possible zero modes dressings define monopoles with the same scaling dimension but distinct quantum numbers. 
Which  type of monopole proliferates is determined by the type of condensed fermion bilinear  and its  spontaneously chosen direction.   

\begin{figure}[ht!]
\centering
{\includegraphics[width = 0.8\linewidth]{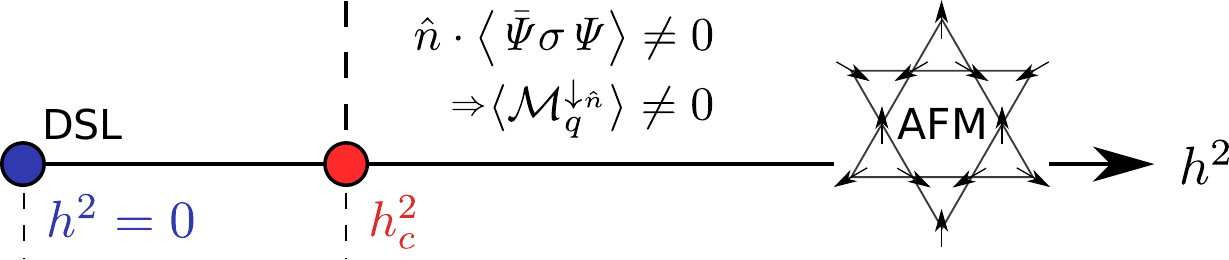}}
\caption{Phase transition from a Dirac spin liquid  to a coplanar antiferromagnet. A condensed spin-Hall mass lets a monopole with spin quantum numbers proliferate.  \label{fig:DSL_to_AFM}}
\end{figure}

In particular, by tuning a chiral-Heisenberg Gross-Neveu ($\cHGN$) interaction,
\eqn{
  \lag_{\QEDtcHGN}  =  \lag_{\mQED_3}  -\frac{h^2}{2} (\Psib \bm{\sigma} \Psi)^2\,,
  }  
 a spin-Hall mass $\hat n \cdot \moye{\Psib \bm{\sigma} \Psi}$ is condensed, and a monopole with non-zero magnetic spin $\man_q^{\d_{\hat n}}$ proliferates. Here, the $3$-vector of Pauli matrices $\bm \sigma$ acts on the magnetic spin. Coming back to the    spin-1/2 kagome Heisenberg antiferromagnet, this model  describes the transition from a DSL to a coplanar antiferromagnet \cite{hermele_properties_2008, lu_unification_2017} as shown in Fig.~\ref{fig:DSL_to_AFM}.  To characterize this $\QEDtcHGN$ quantum critical point (QCP), we obtain the scaling dimension of monopoles.

\section{Scaling dimension of a monopole operator \label{sec:scaling}}
  
We first warm up by examining the non-compact theory and establish the existence of the QCP. To do so,  an auxiliary vectorial boson $\bm \phi = (\phi_1, \phi_2, \phi_3)$ is introduced to decouple the quartic interaction,  ${\lag' = - \Psib \lb \sl{D}_a + \bm{\phi} \cdot \bm{\sigma}  \rb \Psi -  \frac{1}{2 h^2} \bm{\phi}^2}$. The fermions are then integrated out, yielding the following effective action
\eqn{
S_{\rm eff} =   - \nf \ln \det\lb \sl{D}_a + \bm \phi \cdot \bm\sigma \rb  +   \frac{1}{2h^2} \int d^3 x \,   \bm{\phi}^2\,.
} 
In the $\nf=\infty$ saddle point solution, the gauge field vanishes due to gauge invariance $\moye{a_\mu} = 0$, and we take a homogeneous ansatz for the boson $\moye{\bm \phi} = M \hat n$. By solving the resulting gap equation $\pa S_{\rm eff} / \pa M = 0$, we find that the order parameter $\moye{\bm{\phi}}$ condenses for $h^{-2} < h^{-2}_c = 0$.\footnote{Divergences in the gap equation are treated with a Zeta regularization.} The effective action at the critical point is then 
 \eqn{
S^c_{\rm eff} =   - \nf \ln \det\lb \sl{D}_a + \bm \phi \cdot \bm\sigma \rb \,.
\label{eq:Seff}
}

We now proceed to compute the scaling dimension of monopole operators in $\QEDtcHGN$.  Following similar work done in $\QEDt$\cite{borokhov_topological_2003} and in the $CP^{N-1}$ model \cite{metlitski_monopoles_2008}, we employ the state-operator correspondence: An operator $\O$ in a conformal field  theory on $\R^3$ can be mapped to a state $\ket{\O}$ in an alternate theory on  $S^{2} \times \R$. 
The two spacetimes are related by a conformal transformation
\eqn{
\dd r^2 + r^2 \dd \Omega^2 \to 
\dd \tau^2+ R^2 \dd \Omega^{2}\,,  \quad  r=R e^{\tau/R}\,,
}
where the radius of the sphere is $R$. 
With this transformation, the dilatation operator on $\R^3$ is mapped to the Hamiltonian on $S^2 \times \R$. This implies that the scaling dimension of the operator $\O$ is equal to the energy  of its related state, $\Delta_{\O} = E_{\ket{\O}}$. 

For  a monopole operator $\O = \man_q^\dag$, the alternate theory  is obtained by adding  a monopole background gauge field $A^q$ \cite{borokhov_topological_2003} and by putting the effective action~\eqref{eq:Seff} on $S^2 \times \R$. This selects the topological sector of operators with charge $q$.  We  restrain our study to the monopole operator with the minimum scaling dimension $\Delta_q \equiv \min(\Delta_{\man_q})$, which corresponds to the ground state of this alternate theory. But the ground state energy is the free energy. The scaling dimension at leading order in $1/N$,  $\Delta_q = \nf \Delta^{(0)}_q + \O(1/\nf^0)$,  is then given by  the saddle point value of the effective action
\eqn{
\Delta_{q}^{(0)} = \dfrac{1}{\nf} S^c_{\rm eff}[A^q]\Big |_{\moye{a_\mu} = 0\,, \moye{\bm \phi} = M_q  \hat z} = - \ln \Det \lb \sl{D}^{S^2 \times \R}_{A^q} + M_q \sigma_z \rb\,, \label{eq:Seff_prime}
}
where $\moye{\bm{\phi}}$ is again taken homogeneous and along $\hat z$ without loss of generality.

\begin{figure}[ht!]
\centering
\subfigure[\label{fig:M_down}]
{
\begin{tabular}{c}
\\[-12.5em]
\includegraphics[width=0.4\linewidth]{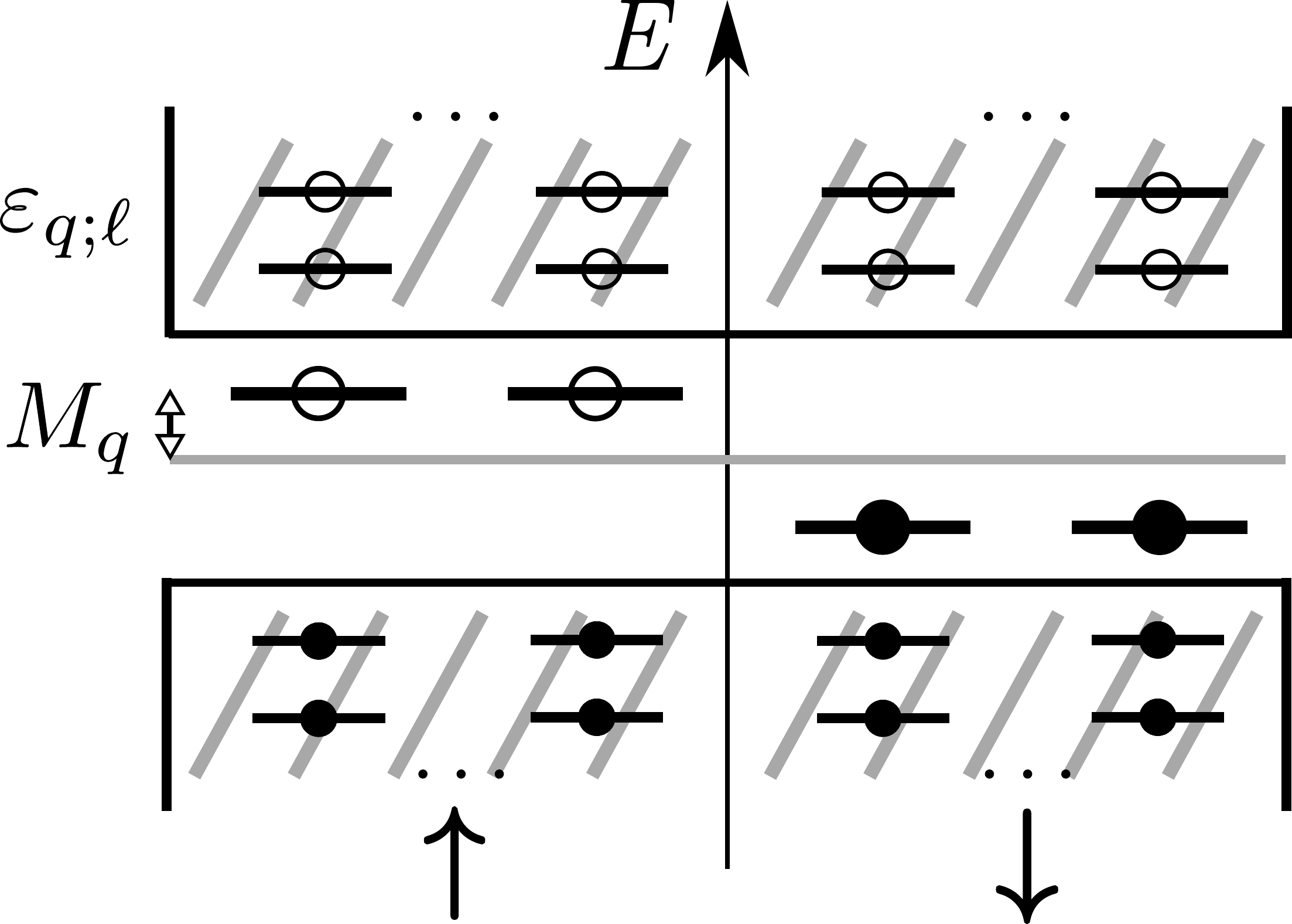}
\end{tabular}
\hspace{-0.25em}
}
\qquad 
\subfigure[\label{fig:mass_scaling}]
{{\includegraphics[width=0.45\linewidth]{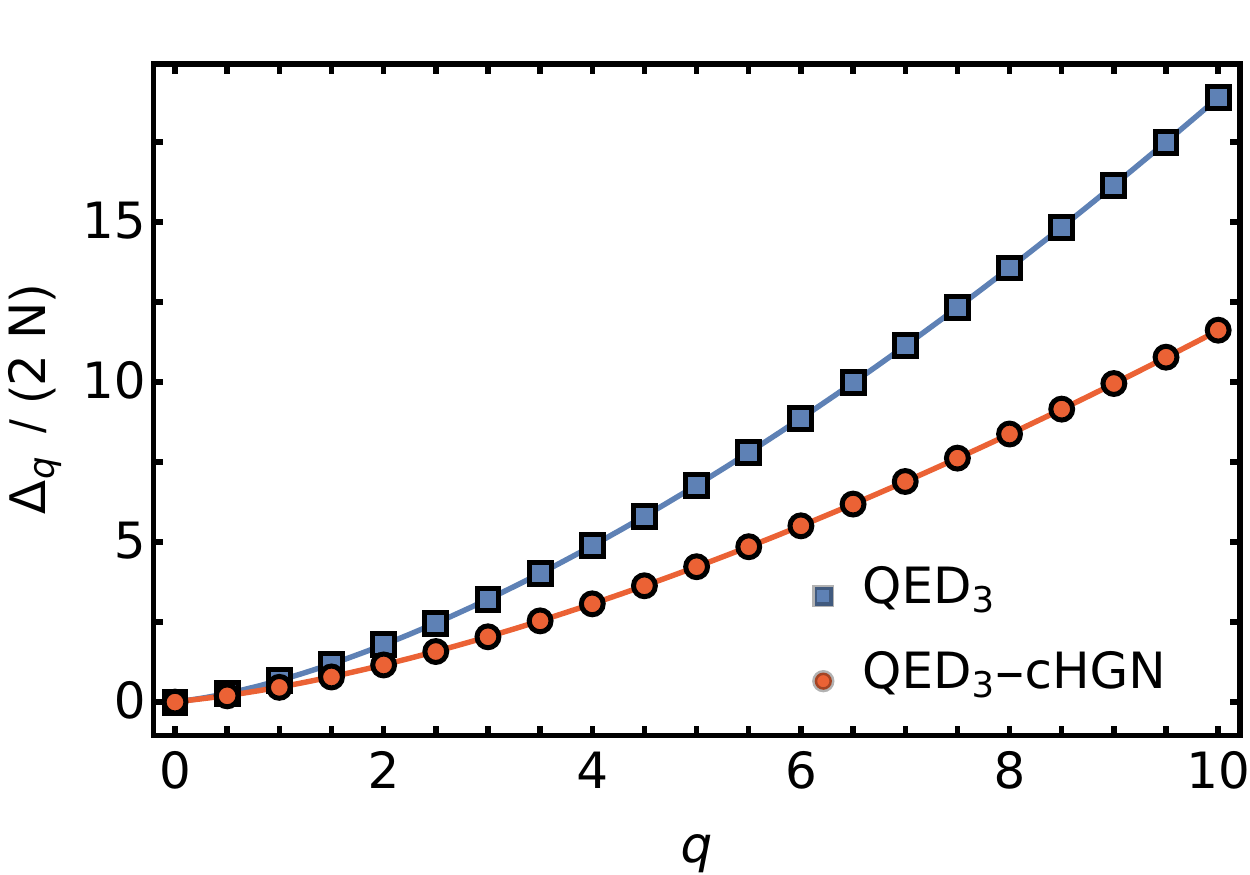}}}
\caption{(a) Schematic representation of the fermion occupation of the monopole operator with the lowest scaling dimension and  with a  spin-Hall mass $\moye{\Psib \bm \sigma \Psi} \propto \moye{\bm \phi} = M_q  \hat z$.  (b) The scaling dimension as a function of the magnetic charge $q$ in $\QEDtcHGN$ and in $\QEDt$ ($M_q = 0$) with the analytical approximation from the large-$q$ expansion.  This figure appears with a shorter range in $q$ in \cite{dupuis_transition_2019}.  \label{fig:}}
\end{figure}

The Dirac operator in Eq.~\eqref{eq:Seff_prime} depends  on generalized  angular momentum $\bm L_q =  \bm{r} \cross \lb \bm p +\bm A^q \rb - q \hat{r} $ and  total spin $\bm J_q = \bm L_q + \bm \tau / 2 $ where $\bm \tau$ acts on particle-hole (Lorentz) space. Spinor monopole harmonics $S^\pm_{q;\ell,m}$ diagonalize  $L_q^2,  J_q^z$, and ${J_q^2 \to j_{\pm} \lb j_{\pm}+1 \rb}$ where $j_{\pm} = \ell \pm 1/2$.  These are two-component spinors built with generalized spherical harmonics $Y_{q;\ell m}$ such that $L_q^2 \to  \ell \lb \ell+1 \rb $ and  ${L_q^z \to  m}$, where $\ell = |q|, |q| + 1, \dots$ \cite{wu_dirac_1976}.  Working in the $j=\ell-1/2$ basis $(S^-_{q;\ell,m} , S^+_{q;\ell-1,m} )^\intercal$, the Dirac operator  reduces   to a matrix with c-number entries \cite{borokhov_topological_2003}. As for the spin-Hall mass $M_q  \sigma_z$, it is already diagonal in spacetime and particle-hole space.  The operator in the determinant of Eq.~\eqref{eq:Seff_prime}  then yields the following spectrum~\cite{dupuis_transition_2019}  
\eqn{
\omega  + i \sigma  M_q & \,,     \quad \ell=q  \,, 
\label{eq:spectrum-a}
\\
\pm \sqrt{\omega^2 + \veps_\ell^2} & \,, \quad \ell= \{q +1, q +2, \dots\}  \,,
\label{eq:spectrum-b}
}
where $\veps_{\ell} = R^{-1}\sqrt{\ell^2 -q^2 + M_q^2 R^2}$ is the energy, $\sigma = \pm 1$ is the spin projection and the magnetic charge is taken positive $q>0$. 
Importantly, the zero modes of $\QEDt$ corresponding to a minimum angular momentum $\ell = q$ now have a non-zero energy, positive for spin up modes and negative for spin down modes. As monopole operators are still dressed with half of those ``zero" modes, the minimal scaling dimension is now obtained by filling only the spin down ``zero" modes. The spectrum \eqref{eq:spectrum-a}-\eqref{eq:spectrum-b} is inserted in Eq.~\eqref{eq:Seff_prime} to obtain the scaling dimension
\eqn{
\Delta_q   
&= - \nf \bigg(  d_q M_q + \sum_{\ell = q  +1}^\infty 2 d_\ell \veps_\ell \bigg)  + \O(1/\nf^0)  \,, \quad d_{\ell} = 2\ell\,,
\label{eq:delta_unreg}
}
where   the radius $R$ was eliminated with rescaling. This indeed corresponds to the energy obtained by filling  all valence modes and spin down ``zero" modes as represented in Fig.~\ref{fig:M_down}.

The gap equation  ${\pa \Delta^{(0)} / \pa M_q = 0}$ can  be solved for the non-trivial expectation value of the spin-Hall mass ${\moye{|\bm \phi|} = M_q}$\,. We stress that this mass~$M_q$ defines a monopole operator at the QCP and is not an indication of spontaneous symmetry breaking in the model.    The scaling dimension is computed by inserting $M_q$ in a regularized version of Eq.~\eqref{eq:delta_unreg}. The  spin-Hall mass $M_q$ and the monopole operator scaling dimension $\Delta_q$ obtained numerically are shown for a few magnetic charges in Fig.~\ref{fig:mass_scaling}. The case of $\QEDt$, where there is no fermion self-interaction and $M_q=0$,  is also shown.  Full lines in Fig.~\ref{fig:mass_scaling} are analytical approximations obtained for a large magnetic charge $q$. 
Note that the scaling dimension $\Delta_q$ is smaller in $\QEDtcHGN$ than in $\QEDt$. For the minimal magnetic charge $q=1/2$, the scaling dimension at the QCP is 
\eqn{
\Delta_{q = 1/2} =   2 \nf \times 0.195 + \O(1/\nf^0)\,.
\label{eq:scaling}
}
For $2\nf =2.56$, this seems to indicate a unitary bound violation, $\Delta_{q=1/2} < 1/2$.

\section{Hierarchy among monopole operators}
So far, we have only computed the minimal scaling dimension of monopole operators in $\QEDtcHGN$. Monopole operators with larger scaling dimensions are expected at this QCP since the degeneracy of monopoles in $\QEDt$ should be lifted~\cite{dupuis_transition_2019} by the $\cHGN$ interaction which breaks the flavor symmetry
\eqn{
 \SU(2 \nf)  \to \SU(2) \times \SU(N)\,.
 \label{eq:symmetry_breaking}
 } 
We first review how monopole operators are organized in  $\QEDt$ with flavor symmetry $\SU(2 \nf)$. We focus on the simplest case with the minimal magnetic charge ${q=1/2}$ where monopole operators are automatically Lorentz scalars \cite{borokhov_topological_2003}. A monopole operator can then be written as  half of the $2 \nf$ zero modes creation operators $c^{\dag}_{I_{i}}$ multiplying a bare monopole operator $\man_{\rm Bare}^\dag$ which creates all negative energy modes in a $2\pi$ flux background 
\eqn{
\man^\dag_{I_1 \dots I_N} = c^\dag_{I_1} \dots c^{\dag}_{I_{N}} \man^\dag_{\rm Bare}\,, \quad I_i \in \{1, 2, \dots, 2 \nf\}\,.}
Antisymmetry between these fermionic operators yields a rank-$\nf$  antisymmetric tensor in flavor space. The first step in understanding the hierarchy of monopole operators  at the QCP is to obtain the reduction of this  $\SU(2\nf)$ irreducible representation (irrep) in terms of irreps of the subgroup  $\SU(2) \times \SU(\nf)$. This is a specific case of the  reduction  $\SU(M N) \to  \SU(M) \times \SU(N)$,  where $M$ and $N$ are integers, whose branching rules are well studied~\cite{itzykson_unitary_1966}.

\begin{figure}[h!]
\centering
\subfigure[\label{fig:}]
{\includegraphics[height=1.5 cm]{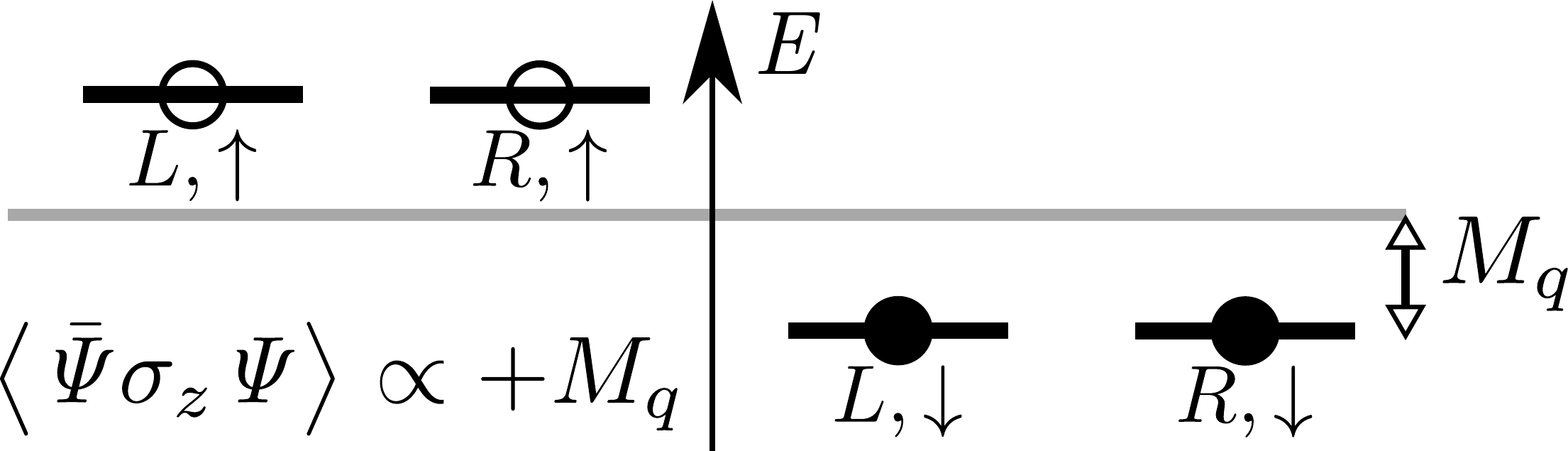}}
\qquad 
\subfigure[\label{fig:}]
{\includegraphics[height=1.5 cm]{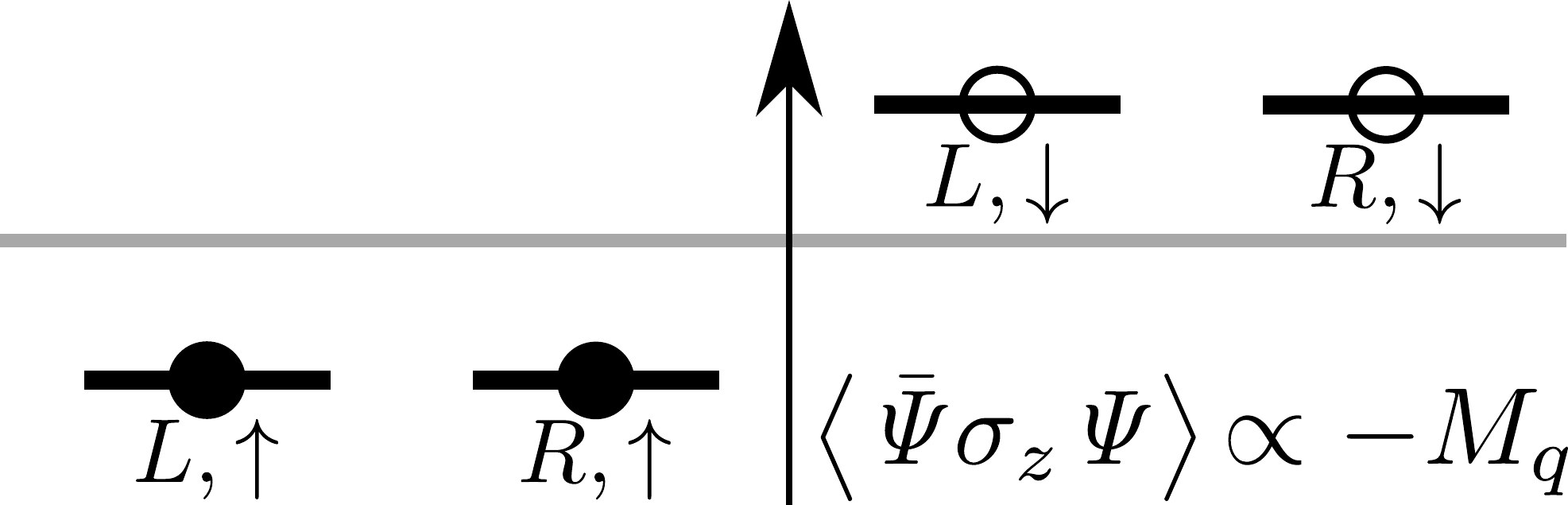}}
\caption{Fermion zero modes occupation of monopole operators for $\nf =2$ and $q=1/2$. (a)  The spin up monopole  and (b) the spin down monopole  are related by a $\SU(2)_{\rm spin}$ rotation which changes the sign of the spin-Hall mass. Here, $M_q>0$. \label{fig:spin_flip}}
\end{figure}

For $N = 2$ , there are two valleys  $v = L, R$ and monopoles form the rank-$2$ antisymmetric irrep of $\SU(4)$, denoted by its dimension $\bm 6$. We note that for the specific examples discussed in this section, the irreps are uniquely determined by their dimension. Monopole operators are then expressed as
\eqn{
c^\dag A(c^{\dag})^{\intercal}  \man^\dag_{\rm Bare}\,,
}
where $A$ acts on vectors in  flavor space  $c^\dag =  \big( c_{\u, L}^\dag,\,  c_{\d, L}^\dag,\,  c_{\u, R}^\dag,\,  c_{\d, R}^\dag \big)$. 
At the QCP, monopole operators in the  irrep $\bm 6$ of $\SU(4)$ of $\QEDt$  reorganize as irreps $(\bm m, \bm n)$ with dimension $m \times n$ of the remaining subgroup ${\SU(2)_{\rm Spin} \times \SU(2)_{\rm Nodal}}$~\cite{itzykson_unitary_1966}
\eqn{
\bm 6 \to \lb \bm 3, \bm 1\rb \oplus \lb \bm 1 , \bm 3 \rb\,.
}
Monopoles are then decomposed as  spin and nodal triplets, respectively $\lb \bm 3, \bm 1\rb$ and $\lb \bm 1, \bm 3\rb$, which may be written as~\cite{hermele_properties_2008}
\eqn{
\vec{\man}_{\rm Spin}^{\dag }  = c^\dag \lb \sigma_y \bm \sigma  \otimes  \mu_y \rb (c^{\dag})^{\intercal} \, \man_{\rm Bare}^{\dag}\,, \quad 
\vec{\man}_{\rm Nodal}^{\dag }  = c^\dag \lb \sigma_y \otimes  \mu_y \bm \mu \rb (c^{\dag})^{\intercal} \, \man_{\rm Bare}^{\dag}\,,
}
where $\bm \mu$ and Pauli matrices acting on nodal subspace. 
The scaling dimension of monopole operators in these spin and nodal triplets are expected to differ  given that no hidden symmetry connects the multiplets. Monopole operators with the largest total spin (here, $S=1$) have access to the largest polarization and as a result can minimize the contribution from the spin-Hall mass. These operators have the lowest scaling dimension and are the finite-$\nf$ analogues of monopoles with minimal scaling dimension $\Delta_{1/2}$ studied in Sec.~\ref{sec:scaling}. In our large-$\nf$ analysis, we gave the example of a monopole filled only with spin down ``zero" modes along $\hat z$ such that the corresponding spin-Hall mass ${\moye{\Psib \sigma_z \Psi} \propto M_q > 0}$ is minimized. As we perform a $\SU(2)_{\rm Spin}$ transformation rotating the  spin down modes to spin up modes, the spin-Hall mass sign changes  ${\moye{\Psib \sigma_z \Psi} \to - \moye{\Psib \sigma_z \Psi}}$.  This leaves the scaling dimension unchanged as we rotate to another monopole of the triplet as schematically shown in Fig.~\ref{fig:spin_flip}. 
A rotation could also make the spin-Hall mass point in the $\hat x$ direction, in which case a combination of all the states in the spin triplet would yield the monopole operator polarized along $\hat x$. Note that this specific element was not well addressed in \cite{dupuis_transition_2019} as the possibility of a rotating the spin-Hall mass was not discussed.

Further insight is obtained by considering the QCP with larger values of $\nf$.  For $\nf=3$, the flavor symmetry is reduced as ${\SU(6) \to \SU(2)_{\rm Spin} \times \SU(3)_{\rm Nodal}}$ at the QCP, and the rank-$3$ antisymmetric representation  $\bm{20}$ of $\SU(6)$ decomposes as  \cite{itzykson_unitary_1966}
\eqn{
\bm{20} \to \lb \bm 4, \bm 1 \rb \oplus \lb \bm 2, \bm 8 \rb\,.
\label{eq:20}
}
The RHS has dimension $4 \times 1 + 2 \times 8 = 20$ as required. 
Again,  monopole operators with the highest $\SU(2)_{\rm spin}$ spin, here a spin quadruplet $S=3/2$, the $\bm 4$ in Eq.~\eqref{eq:20}, have the lowest scaling dimension. As for the spin doublet $S=1/2$ (denoted $\bm 2$), it is obtained by the composition of three spins : Two spins form a singlet while the remaining spin is either up or down.   For general $\nf$, monopole operators reorganize  as various  magnetic spin multiplets with total spin $S_{\rm min} \leq S \leq (\nf + 1 )/2 $, with minimum spin  $S_{\min} = 0$  for $\nf$ even and $S_{\min} = 1/2$  for $\nf$ odd. For large-$\nf$, this distinction does not affect the scaling dimension.  The almost equally populated spin up and spin down ``zero" modes of a spin doublet $S=1/2$ monopole mostly cancel their contribution to the scaling dimension. The remaining unpaired ``zero" mode has a contribution order $\O(1/\nf^0)$  which is neglected in our leading order treatment. Therefore, the largest scaling dimension (characterizing monopoles with either $S =0$ or $S = 1/2$) is given, at leading order in $1/\nf$, by the second term  in $\eqref{eq:delta_unreg}$.
The gap equation for this expression is solved for $M_q = 0$, which yields the same scaling dimension as in $\QEDt$. Monopole operators in $\QEDtcHGN$ then have scaling dimensions $\Delta_{\man_q}$ which vary between the lowest monopole scaling dimension in $\QEDtcHGN$ and $\QEDt$, 
\eqn{
\Delta_q^{\QEDtcHGN} \leq \Delta_{\man_q}^{\QEDtcHGN} \leq \Delta_q^{\QEDt}\,, \quad \text{leading order in } 1/\nf\,.
}
This upper boundary was overestimated in \cite{dupuis_transition_2019} as the possibility of  different masses $M_q$ defining each monopole multiplet was not considered.

\paragraph{Conclusion -}
The DSL is a parent state of many quantum phases and exotic non-Landau transitions in frustrated quantum magnets. We characterized monopole operators in $\QEDtGN$ models describing various QCPs between this DSL and other ordered phases. Notably, we obtained the lowest scaling dimension of monopole operators in  the $\QEDtcHGN$ model.  At leading order in $1/\nf$, we find it is lower than  its counterpart in $\QEDt$. Specifically, for a minimal magnetic charge, it is given by  $\Delta_{q=1/2}^{\QEDtcHGN} = 2 \nf \times 0.195 + \O(1/\nf^0)$. We discussed how monopoles are reorganized as irreps of the $\QEDtcHGN$ QCP reduced symmetry group $\SU(2) \times \SU(\nf)$.  Monopoles multiplets with the highest $\SU(2)$ spin have the lowest scaling dimension.     A more detailed exploration of this monopole hierarchy at the QCP  is reserved for future work. 

\paragraph{Acknowledgements -} 
  We thank  Tarun Grover, Joseph Maciejko, David Poland,  Zi Yang Meng, Sergue\"i Tchoumakov and  Chong Wang for the interesting discussions and insightful observations on our work  during the symposium. {\'E.D.} was funded by an Alexander Graham Bell CGS from NSERC. {M.B.P.} was funded by a Discovery Grant from NSERC. {W.W.-K.} was funded by a Discovery Grant from NSERC, a Canada Research Chair, a grant from the Fondation Courtois, and a ``\'Etablissement de nouveaux chercheurs et de nouvelles chercheuses universitaires" grant from FRQNT.

%

%
\end{document}